\documentclass[a4paper,fleqn]{cas-sc}
\usepackage[numbers]{natbib}
\usepackage{hyperref}
\usepackage[dvipsnames]{xcolor}

\def\tsc#1{\csdef{#1}{\textsc{\lowercase{#1}}\xspace}}
\tsc{WGM}
\tsc{QE}
\tsc{EP}
\tsc{PMS}
\tsc{BEC}
\tsc{DE}

\begin{document}
\let\WriteBookmarks\relax
\def\floatpagepagefraction{1}
\def\textpagefraction{.001}

\shorttitle{Protecting three-dimensional entanglement from correlated amplitude damping channel}

\shortauthors{X.~Xiao, W.~R.~Huang, T.~X.~Lu, and Y.~L.~Li}

\title[mode = title]{Protecting three-dimensional entanglement from correlated amplitude damping channel}    
\author[1]{Xing Xiao}[orcid= 0000-0001-6925-5923]
\author[1]{Wen-Rui Huang}[orcid=0000-0002-4781-4090]
\author[1]{Tian-Xiang Lu}[orcid=0000-0002-4781-4090]

\author[2]{Yan-Ling Li}[orcid=0000-0002-7807-1496]
\ead{liyanling0423@gmail.com}
\cormark[1]
\cortext[cor1]{Corresponding author}

\affiliation[1]{organization={College of Physics and Electronic Information, Gannan Normal University},
    city={Ganzhou, Jiangxi},
    postcode={341000}, 
    country={China}}
\affiliation[2]{organization={School of Information Engineering, Jiangxi University of Science and Technology},
     city={Ganzhou, Jiangxi},
    postcode={341000}, 
    country={China}}

\begin{abstract}
Quantum entanglement is a crucial resource in quantum information processing, and protecting it against noise poses a significant challenge. This paper introduces two strategies for preserving qutrit-qutrit entanglement in the presence of correlated amplitude damping (CAD) noise: weak measurement (WM) and environment-assisted measurement (EAM), both combined with quantum measurement reversal (QMR). Two prototypical classes of three-dimensional entangled states are examined. The findings demonstrate that while the WM+QMR method can partially retain entanglement, the EAM+QMR approach is more effective at protecting entanglement as well as enhancing success probabilities, particularly for specific qutrit-qutrit entangled states. Additionally, we thoroughly discuss the impact of correlation effects on entanglement protection and the enhancement of success probability. Our results provide valuable insights into defending high-dimensional entanglement from CAD noise, thus offering practical solutions for the advancement of quantum information technologies.
\end{abstract}




\begin{keywords}
quantum entanglement \sep correlated amplitude damping noise \sep weak measurement \sep environment-assisted measurement.
\end{keywords}

\maketitle

\section{Introduction}
\label{sec1}
Quantum entanglement represents one of the most fundamental and intriguing phenomena in the field of quantum physics. It describes the nonlocal correlations that exist between two or more quantum systems. Entanglement plays a pivotal role in quantum information processing \cite{nielsen_2019_quantum}, as evidenced by the numerous applications of this phenomenon, including quantum cryptography, quantum teleportation, quantum computation, and quantum metrology. Nevertheless, the majority of existing quantum information protocols are based on qubits. Although the qubit (two-level system) is the basic unit of quantum information, it has limited information capacity and robustness against noise. Consequently, there is a growing interest in exploring high-dimensional quantum systems, which include the qutrit (three-level system) and the qudit ($d$-level system). In fact, due to the extra dimensions of the Hilbert space, high-dimensional entanglement has been shown to have many advantages over qubit entanglement, such as higher channel capacity for quantum communication \cite{barreiro_2008_beating,pirandola_2017_fundamental,miller_2019_parameter}, higher fidelity for quantum teleportation, higher precision for quantum metrology \cite{fickler_2012_quantum} and higher error tolerance for noisy quantum information tasks \cite{cerf_2002_security,ecker_2019_overcoming}. These advantages are particularly intriguing in the context of the upcoming noisy intermediate-scale quantum (NISQ) era, given that the ``quality'' is of critical importance \cite{preskill_2018_quantum}. In recent years, quantum information processing based on qutrit has made great progress in experiments, for example, entanglement generation \cite{erhard_2020_advances,hu_2020_efficient,cerveralierta_2022_experimental}, three-dimensional quantum teleportation \cite{luo_2019_quantum,hu_2020_experimental}, quantum superdense coding \cite{hu_2018_beating}, two-qutrit quantum gate \cite{luo_2023_experimental,goss_2022_highfidelity}, and so on.

As is well known, quantum entanglement is susceptible to decay due to environmental noise and is even associated with the phenomenon of entanglement sudden death. This is also true of high-dimensional quantum entanglement. 
In recent decades, there has been a growing interest in the study of noise in quantum entanglement. Many methods have been developed to protect the quantum entanglement from the environmental noises, such as quantum error correction, dynamic decoupling and quantum feedback control. Each method has its own advantages and disadvantages, and the choice of method depends on the specific type of noise (e.g., dissipative or dephasing) and the physical system being used.
However, the majority of previous investigations have concentrated on the uncorrelated noise, where the noises are assumed to have different origins and are treated independently. While this presumption is reasonable in specific physical scenarios, it may not be justifiable in numerous more practical circumstances \cite{macchiavello_2002_entanglementenhanced,banaszek_2004_experimental,darrigo_2007_quantum,darrigo_2013_classical}. The noises may share a common origin, resulting in their correlation with one another. This makes the noise effects not independent \cite{uvonlpke_2020_twoqubit}. For instance, successive use of the same channel may cause the correlated noise between inputs, as the channel may retain memory between successive transmissions. Therefore, the study of correlated noise is of significant importance for the understanding and controlling the performance of quantum information processes. Moreover, some researchers have identified some beneficial effects of correlated noise in quantum information processes. These include the protection of quantum coherence \cite{lan_2023_protecting}, the realization of quantum teleportation \cite{sun_2022_memory}, and the enhancement of quantum Fisher information \cite{li_2023_enhancing}.
Most of the previous studies have been confined to the two-dimensional Hilbert space. In a recent work, the authors show that the correlated channel can generally enhance the robustness of qutrit teleportation to noise \cite{xu_2022_enhancing}. However, the results become invalid in severe noise regions. Such an invalidity can be attributed to the degradation of initially shared qutrit-qutrit 
entanglement since the state fidelity in the standard teleportation protocol is directly proportional to the entanglement. Therefore, the protection of high-dimensional quantum entanglement is a crucial matter, as it determines whether the advantages of high-dimensional quantum systems can be realized. It is of significant interest to determine the optimal approach for leveraging the correlation effect in the presence of correlated noise, while simultaneously minimizing its detrimental impact.

This paper presents two methods that protect the qutrit-qutrit entanglement, despite the presence of correlated amplitude damping (CAD) noise. We achieve this by employing the techniques of weak measurement (WM) and environment-assisted measurement (EAM). WM is a form of measurement that is designed to avoid the complete collapse of the wave function \cite{korotkov_2006_undoing,katz_2008_reversal,korotkov_2010_decoherence,lee_2011_experimental}. By choosing the proper post QMR, the effects of CAD noise and WM can be eliminated, thereby restoring the original entanglement. The effectiveness of WM in protecting entanglement from AD noise has been demonstrated in various theoretical and experimental studies \cite{kim_2011_protecting,li_2013_recovering,man_2012_manipulating,wang_2014_protecting,xiao_2016_protecting,xiao_2018_retrieving,xiao_2020_enhanced,foletto_2020_experimental,harraz_2022_enhancing}. However, its ability to manage CAD noise, particularly qutrit CAD noise, has yet to be investigated. In the EAM approach, a measurement, such as photon counting, is carried out on the noisy environment coupled to the qutrits \cite{zhao_2013_restoration,wang_2014_environmentassisted,goldblatt_2024_recovering}. This is followed by a QMR operation on the system, which is conditioned by the measurement results. The findings of our study indicate that the utilization of EAM facilitates the near-perfect recovery of the initial entanglement, thereby eliminating the decoherence effects of the CAD noise.

The following section outlines the structure of the paper. In Sec. \ref{sec2}, we introduce the theoretical model of CAD noise in qutrit-qutrit system, derive the Kraus operators of AD noise and FCAD noise, and provide the dynamical map of CAD noise. In Sec. \ref{sec3}, we elucidate the behavior of entanglement in CAD noise with two types of initial entangled states. In Sec. \ref{sec4}, the protection of the qutrit-qutrit entanglement under CAD noise by WM and QMR is investigated. Sec. \ref{sec5} demonstrates the second approach of improving the entanglement in a CAD noisy channel by using the combination of EAM and QMR, and provides a brief comparison between these two schemes. In Sec. \ref{sec6}, we discuss the experimental feasibility of our procedures. Sec. \ref{sec7} is devoted to the conclusions.
\section{CAD noise in qutrit-qutrit system}
\label{sec2}
The present study considers two $V$-type three-level systems, with levels $|0\rangle$, $|1\rangle$ and $|2\rangle$, which are successively interacted with a zero-temperature bath. In the simplest scenario, the time interval between two qutrits passing through the channel is sufficiently long, which indicates the consecutive uses of the bath have no correlations. In this case the dynamical map will be restricted to be the tensor product of the maps of the individual qutrits $\mathcal{E}_{n}=\mathcal{E}^{\otimes n}$. In the Born-Markov approximation, the map of each qutrit can be described by the Lindblad equation
\begin{equation}
\label{eq1}
\dot{\rho}=\frac{\gamma_{1}}{2}(2\sigma_{01}\rho\sigma_{10}-\sigma_{11}\rho-\rho\sigma_{11})+\frac{\gamma_{2}}{2}(2\sigma_{02}\rho\sigma_{20}-\sigma_{22}\rho-\rho\sigma_{22}),
\end{equation}
where $\gamma_{1}$ and $\gamma_{2}$ are the spontaneous rates of the excited levels $|1\rangle$ and $|2\rangle$, respectively. $\rho\in C^{3\times3}$ is the density matrix of the qutrit. $\sigma_{ij}=|i\rangle\langle j|$ is a transition operator between $|i\rangle$ and $|j\rangle$. According to the operator-sum representation $\rho(t)=\sum_{i}E_{i}\rho E_{i}^{\dagger}$, one can find the Kraus operators
\begin{equation}
\label{eq2}
E_{0}=\left(\begin{array}{ccc}1 & 0 & 0 \\0 & \sqrt{1-{d}_1} & 0 \\0 & 0 & \sqrt{1-{d}_2}\end{array}\right),E_{1}=\left(\begin{array}{ccc}0 & \sqrt{d_1} & 0 \\0 & 0 & 0 \\0 & 0 & 0\end{array}\right),
E_{2}=\left(\begin{array}{ccc}0 & 0 & \sqrt{d_2} \\0 & 0 & 0 \\0 & 0 & 0\end{array}\right),
\end{equation}
with $d_1=1-\exp(-\gamma_{1}t)$, $d_2=1-\exp(-\gamma_{2}t)$. It's natural to obtain the Kraus operators of two-qutrit uncorrelated AD channel are tensor products of the single-qutrit Kraus operators $E_{ij}=E_{i}\otimes E_{j}$ with $i,j=0,1,2$.

However, if the time interval between two qutrits passing through the channel is extremely short, then the consecutive uses of the bath are almost perfectly correlated. Formally, the evolution can be expressed as the correlated Lindblad equation of the two-qutrit state $\tilde{\rho}$ \cite{yeo_2003_timecorrelated}. 
\begin{equation}
\label{eq3}
\dot{\tilde{\rho}}=\frac{\gamma_{1}}{2}\big(2\sigma_{01}^{\otimes2}\tilde{\rho}\sigma_{10}^{\otimes2}-\sigma_{11}^{\otimes2}\tilde{\rho}-\tilde{\rho}\sigma_{11}^{\otimes2}\big)+\frac{\gamma_{2}}{2}\big(2\sigma_{02}^{\otimes2}\tilde{\rho}\sigma_{20}^{\otimes2}-\sigma_{22}^{\otimes2}\tilde{\rho}-\tilde{\rho}\sigma_{22}^{\otimes2}\big),
\end{equation}
where $\sigma_{ij}^{\otimes2}=\sigma_{ij}\otimes\sigma_{ij}$. Note that $\tilde{\rho}\in C^{9\times9}$ is the total state of two qutrits in the standard basis.  
The Kraus operators of the correlated part of the evolution of two-qutrit state are given as
\begin{equation}
\label{eq4}
A_{00}=\left(\begin{array}{cccc}{\rm \bf I}_{4\times4} & {\rm \bf 0}_{4\times1} & {\rm \bf 0}_{4\times3} & {\rm \bf 0}_{4\times1} \\  {\rm \bf 0}_{1\times4} & \sqrt{1-{d}_1} & {\rm \bf 0}_{1\times3} & {\rm \bf 0}_{1\times1} \\  {\rm \bf 0}_{3\times4} &  {\rm \bf 0}_{3\times1} & {\rm \bf I}_{3\times3} & {\rm \bf 0}_{3\times1}  \\  {\rm \bf 0}_{1\times4} &  {\rm \bf 0}_{1\times1} & {\rm \bf 0}_{1\times3}  & \sqrt{1-{d}_2}\end{array}\right), 
A_{11}=\left(\begin{array}{c|c|c} {\rm \bf 0}_{1\times4}  & \sqrt{d_1} & {\rm \bf 0}_{1\times4}   \\
\hline
 {\rm \bf 0}_{4\times4}   & {\rm \bf 0}_{4\times1}   &  {\rm \bf 0}_{4\times4} \\
\hline
  {\rm \bf 0}_{4\times4} & {\rm \bf 0}_{4\times1}   & {\rm \bf 0}_{4\times4}  \end{array}\right),
A_{22}=\left(\begin{array}{c|c|c} {\rm \bf 0}_{1\times4}  &  {\rm \bf 0}_{1\times4}  & \sqrt{d_2}   \\
\hline
 {\rm \bf 0}_{4\times4}   & {\rm \bf 0}_{4\times4}   &  {\rm \bf 0}_{4\times1} \\
\hline
  {\rm \bf 0}_{4\times4} & {\rm \bf 0}_{4\times4}   & {\rm \bf 0}_{4\times1}  \end{array}\right)
\end{equation}
where ${\rm \bf I}_{n\times n}$ is the $n$-dimensional identity matrix and $ {\rm \bf 0}_{n\times m}$ is the zero matrix of $n\times m$.

\begin{figure}
\centering
\includegraphics[width = 16 cm]{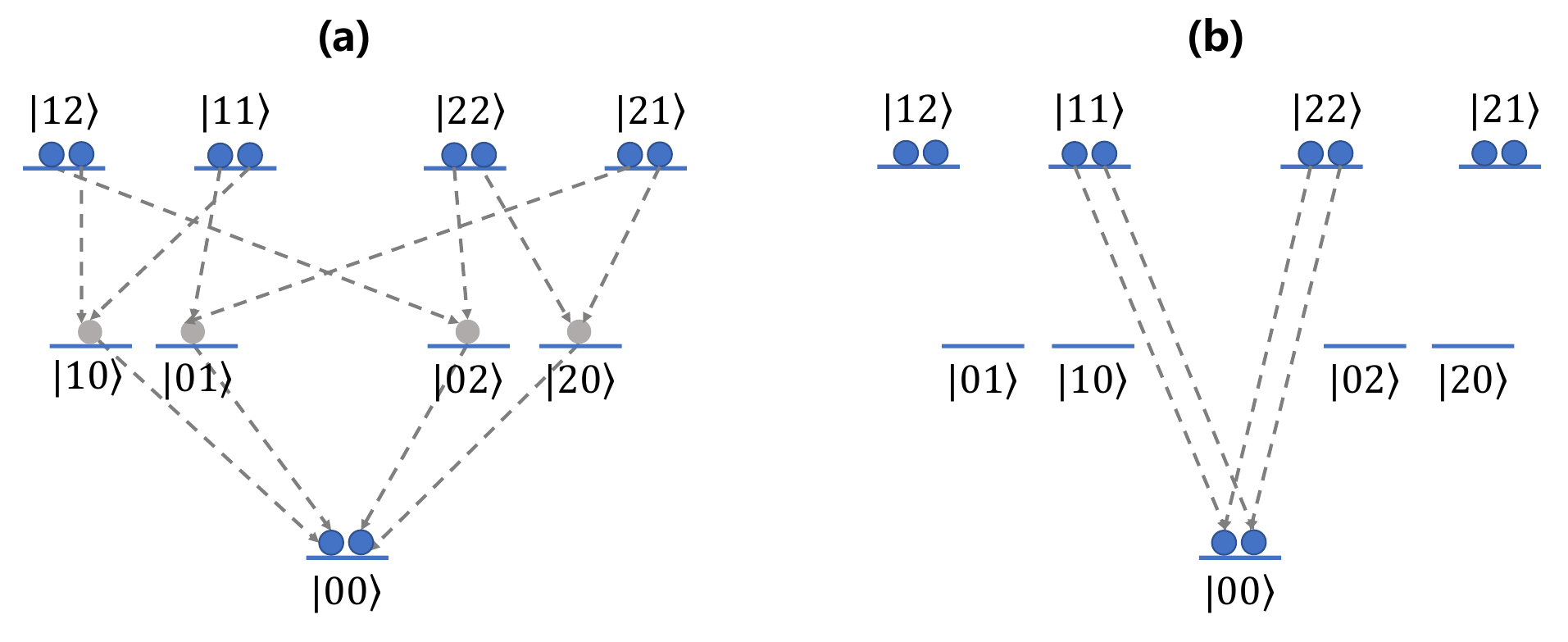}
\caption{\label{fig1} Schematic illustrations of the relaxation mechanism of two $V$-type qutrits. (a) uncorrelated AD noise, (b) fully correlated AD noise.}
\end{figure}

In the general case, the map of two qutrits subjected to CAD channel is a combination of the uncorrelated AD channel and fully correlated AD (FCAD) channel:
\begin{equation}
\label{eq5}
\mathcal{E}_{\rm CAD}\left[\rho(0)\right]=(1-\mu)\mathcal{E}_{\rm AD}^{\otimes2}\left[\rho(0)\right]+\mu\mathcal{E}_{\rm FCAD}\left[\rho(0)\right]
=(1-\mu)\sum_{i,j=0}^{2}E_{ij}\rho(0)E_{ij}^{\dagger}+\mu\sum_{k=0}^{2}A_{kk}\rho(0)A_{kk}^{\dagger},
\end{equation}
where $\mu\in[0,1]$ is the correlation parameter which can be interpreted as the associated probability of fully correlated relaxation.
Figure \ref{fig1} illustrates the relaxation mechanism of the uncorrelated AD channel and the fully correlated AD channel. In Fig. \ref{fig1}(a), the two qutrits decay independently, whereas in Fig. \ref{fig1}(b), there are only synchronized transitions in the fully correlated AD channel.

\section{Qutrit-Qutrit entanglement under CAD channel}
\label{sec3}
A number of entangled qutrit states can be constructed from the basis states of the $H^3\otimes H^3$ Hilbert space. These states can be divided into two groups according to the behaviors of the decay of their entanglement \cite{brnner_2013_robust}. In this paper, we examine two illustrative examples of entangled qutrit states, which are given by
\begin{eqnarray}
\label{eq6}
|\psi\rangle_{1} &=&\alpha |00\rangle +\beta |11\rangle+\gamma |22\rangle,\\
|\psi\rangle_{2} &=&\alpha |02\rangle +\beta |20\rangle+\gamma |11\rangle,
\end{eqnarray}
with $\alpha^{2}+|\beta|^{2}+|\gamma|^{2}=1$. Substituting $\rho_{1}(0)=|\psi\rangle_{1}\langle\psi|$ into equation (\ref{eq5}), we obtain the output state of two qutrits denoted by $\rho_{1}$ after successive passages through the CAD channel.
The non-zero elements of the final state $\rho_{1}$ are
\begin{eqnarray}
\rho_{11} &=& \alpha^{2}+|\beta|^{2}[(1-\mu)d_{1}^{2}-\mu d_{1}]+|\gamma|^{2}[(1-\mu)d_{2}^{2}-\mu d_{2}],\nonumber\\
\rho_{15} &=& \rho_{51}^{*} = \alpha \beta^{*}[(1-\mu)(1-d_{1})+\mu \sqrt{1-d_{1}}],\nonumber\\
\rho_{19} &=& \rho_{91}^{*} = \alpha \gamma^{*}[(1-\mu)(1-d_{2})+\mu \sqrt{1-d_{2}}],\nonumber\\
\rho_{22} &=& \rho_{44} = |\beta|^{2}(1-\mu)(1-d_{1})d_{1},\\
\rho_{33} &=& \rho_{77} = |\gamma|^{2}(1-\mu)(1-d_{2})d_{2},\nonumber\\
\rho_{55} &=& |\beta|^{2}[(1-\mu)(1-d_{1})^{2}+\mu (1-d_{1})],\nonumber\\
\rho_{59} &=& \rho_{95}^{*} = \beta \gamma^{*}[(1-\mu)(1-d_{1})(1-d_{2})+\mu \sqrt{(1-d_{1})(1-d_{2})}],\nonumber\\
\rho_{99} &=& |\gamma|^{2}[(1-\mu)(1-d_{2})^{2}+\mu (1-d_{2})].\nonumber
\end{eqnarray}

\begin{figure}
\centering
\includegraphics[width = 16 cm]{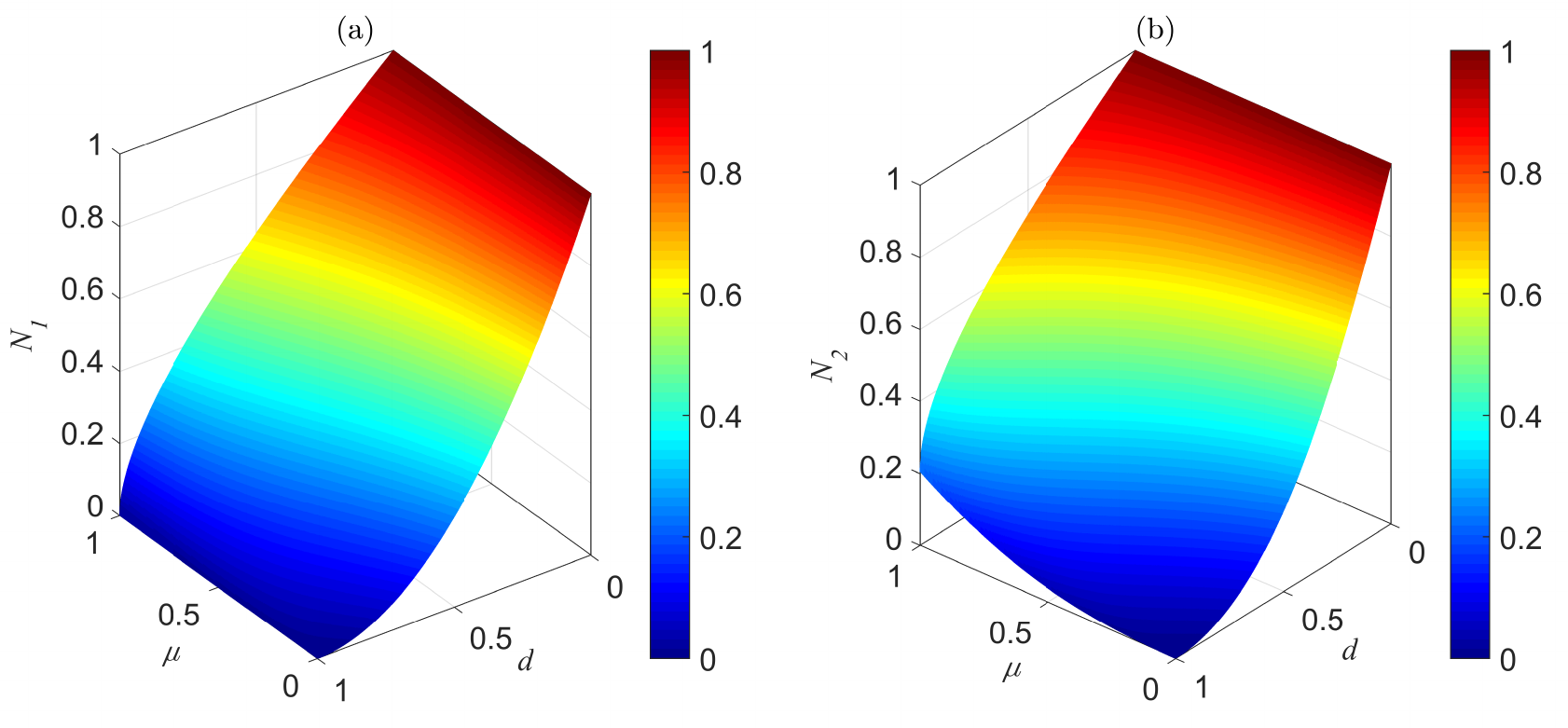}
\caption{\label{fig2} (a) $N_{1}$ and (b) $N_{2}$ as a function of  $d$ and $\mu$. The remaining parameters are set to $\alpha=\beta=\gamma=1/\sqrt{3}$, $d_{1}=d_{2}=d$.}
\end{figure}

In the case of two-dimensional bipartite systems, the concurrence \cite{wootters_1998_entanglement} is a suitable measure of entanglement. However, in the context of mixed states in higher dimensions, the calculation of the concurrence through the construction of a convex roof \cite{mintert_2004_concurrence} is computationally challenging. As an alternative, we choose the negativity as a means of measuring the entanglement of two qutrits \cite{plenio_2005_logarithmic}. The negativity of the state $\rho_{1}$ is
\begin{equation}
N_{1} =\frac{\|\rho_{1}^{T}\|-1}{2},
\label{eq9}
\end{equation}
where $\|\rho_{1}^{T}\|$ represents the trace norm of the partial transposition of the state $\rho_{1}$. The value $N_{1}$ is equal to the absolute value of the sum of negative eigenvalues of $\rho_{1}^{T}$. One can similarly substitute $\rho_{2}(0)=|\psi\rangle_{2}\langle\psi|$ into equation (\ref{eq5}) to obtain the negativity $N_{2}$ of the state $\rho_{2}$. The general analytic expressions of $N_{1}$ and $N_{2}$ are too complicated to present. The behavior of the negativity $N_{1}$ and the negativity $N_{2}$ as a function of the noise parameter and the correlation factor is illustrated in Fig. \ref{fig2}.

In the case of the uncorrelated AD channel, the decay behaviors of the two initial states $|\psi\rangle_{1}$ and $|\psi\rangle_{2}$ are identical. As the noise strength increases, the entanglement gradually decays to zero. However, there is no region of entanglement sudden death \cite{yu_2009_sudden}, which suggests that three-dimensional entanglement is more resilient to noise than two-dimensional quantum entanglement.
Furthermore, Fig. \ref{fig2} illustrates that, despite the initial entanglement being identical, the subsequent evolution differs in the CAD channel. For the initial state $|\psi\rangle_{1}$, it can be observed that the entanglement present within it is completely lost as $d$ approaches 1, regardless of the value of the correlation parameter $\mu$. However, for the initial state $|\psi\rangle_{2}$, it is possible for some entanglement to remain when the noise is correlated. This discrepancy can be attributed to the properties of the initial state and the mechanism of CAD noise. Note that the CAD channel is a combination of the uncorrelated AD channel and the FCAD channel. The uncorrelated AD channel would completely destroy the entanglement because it would allow all excited states to decay to the ground state. While the FCAD channel differs from the AD channel in that it only allows synchronized transitions, i.e. $|11\rangle\rightarrow|00\rangle$ and $|22\rangle\rightarrow|00\rangle$, as shown in Fig. \ref{fig1}. It can be observed that the state $|\psi\rangle_{1}$ is highly susceptible to AD and FCAD channels, whereas the state $|\psi\rangle_{2}$ is comparatively resilient to FCAD channel due to the presence of $|02\rangle$ and $|20\rangle$. Indeed, in the case of some specific states, for instance $|\psi\rangle=\alpha|i,j\rangle+\beta|j,i\rangle+\gamma|k,k\rangle$ with $i\neq j\neq k$ and $i,j,k\in\{0,1,2\}$, it can be found that the presence of the correlation effect makes part of the entanglement immune to the effects of the CAD channel.

%
\section{Protecting qutrit-qutrit entanglement by WM and QMR}
\label{sec4}
In this section, we demonstrate the protection of the qutrit-qutrit entanglement of the CAD channel by WM and QMR. In fact, the utilization of WM and QMR to preserve the entanglement of two initially entangled qutrits, which are subjected to an independent AD channel, was explored in Ref. \cite{xiao_2013_protecting}. Here, we will focus on the CAD channel. A natural question that arises is what would happen if WM and QMR were applied to the case of the CAD channel.

In light of this question, we continue to examine the initially entangled states $|\psi\rangle_{1}$ and $|\psi\rangle_{2}$ and seek to elucidate the role of WM and QMR in protecting entanglement under the two-qutrit CAD channel. Correlated noise may have a positive effect on the preservation of entanglement. However, the results differ from those observed for two-dimensional entanglement. This is due to the fact that three-dimensional entanglement has a richer entanglement structure \cite{xiao_2016_protecting}. 

The process of protecting the entanglement using WM and QMR is divided into three steps. (i) The first step is to perform independent WM on each of the two qutrits. (ii) The two qutrits are then passed through the CAD channel. (iii) Finally, independent QMR is performed on the two qutrits. It should be noted that all of these operations are local in nature, and that no non-local operation is introduced. Consequently, the entanglement does not increase, and at most, it can be restored to its initial value. The process can be expressed as a map
\begin{equation}
\rho_{\rm{WM}} = \frac{1}{\mathcal{N}}M_{\rm R}\left[(1-\mu)\sum_{i,j}E_{ij}\left(M_{\rm WM}\rho(0)M_{\rm WM}^{\dagger}\right)E_{ij}^{\dagger}+\mu\sum_{k}F_{kk}\left(M_{\rm WM}\rho(0)M_{\rm WM}^{\dagger}\right)E_{kk}^{\dagger}\right]M_{\rm R}^{\dagger}.
\label{eq9}
\end{equation}
where $M_{\rm WM} = E_{\rm WM}^{1} \otimes E_{\rm WM}^{2}$, $M_{\rm R} = E_{\rm R}^{1} \otimes E_{\rm R}^{2}$.
As we shall see, WM is related to positive operator-valued measure (POVM). The main advantage of WM is that it is not entirely destructive, thereby allowing for the possibility of restoring the quantum state. For the $V$-type qutrit, the WM and QMR can be expressed as
\begin{eqnarray}
\label{eq8}
E_{\rm WM}=\left(
        \begin{array}{ccc}
          1 & 0
          & 0 \\
          0 & \sqrt{1-p} & 0 \\
          0 & 0 & \sqrt{1-q} \\
        \end{array}
      \right),
      E_{\rm R}=\left(
        \begin{array}{ccc}
         \sqrt{(1-p_{r})(1-q_{r})} & 0
          & 0 \\
          0 & \sqrt{1-q_{r}} & 0 \\
          0 & 0 & \sqrt{1-p_{r}} \\
        \end{array}
      \right).
\end{eqnarray}
where $0 \leq p,q,p_{r},q_{r} \leq 1$ are the strength parameters of WM and QMR.

For the initial state $|\psi\rangle_{1}$, the non-zero elements of the final state $\rho_{1,\rm WM}$ are
\begin{eqnarray}
\rho_{11} &=& \frac{1}{\mathcal{N}}[\alpha^{2}+|\beta|^{2}(\bar{\mu} d_{1}^{2}-\mu d_{1})\bar{p}^{2}
+|\gamma|^{2}(\bar{\mu}d_{2}^{2}-\mu d_{2})\bar{q}^{2}]\bar{p}_{r}^{2}\bar{q}_{r}^{2},\nonumber\\
\rho_{15} &=& \rho_{51}^{*} = \frac{1}{\mathcal{N}}\alpha \beta^{*}(\bar{\mu}\bar{d}_{1}+\mu \sqrt{\bar{d}_{1}})\bar{p} \bar{p}_{r}^{2}\bar{q}_{r},\nonumber\\
\rho_{19} &=& \rho_{91}^{*} = \frac{1}{\mathcal{N}} \alpha \gamma^{*}(\bar{\mu}\bar{d}_{2}+\mu \sqrt{\bar{d}_{2}})\bar{p}_{r}\bar{q} \bar{q}_{r}^{2},\nonumber\\
\rho_{22} &=& \frac{1}{\mathcal{N}}|\beta|^{2}\bar{\mu}\bar{d}_{1}d_{1}\bar{p}^{2} \bar{p}_{r}^{2}\bar{q}_{r},\nonumber\\
\rho_{33} &=& \frac{1}{\mathcal{N}}|\gamma|^{2}\bar{\mu}\bar{d}_{2}d_{2}\bar{p}_{r}\bar{q}_{r}^{2}\bar{q}^{2},\nonumber\\
\rho_{44} &=& \frac{1}{\mathcal{N}}|\beta|^{2}\bar{\mu}\bar{d}_{1}d_{1}\bar{p}^{2} \bar{p}_{r}^{2}\bar{q}_{r},\nonumber\\
\rho_{55} &=& \frac{1}{\mathcal{N}}|\beta|^{2}(\bar{\mu}\bar{d}_{1}^{2}+\mu \bar{d}_{1})\bar{p}^{2}\bar{p}_{r}^{2},\nonumber\\
\rho_{59} &=& \rho_{95}^{*} = \frac{1}{\mathcal{N}}\beta \gamma^{*}(\bar{\mu}\bar{d}_{1}\bar{d}_{2}+\mu \sqrt{\bar{d}_{1}\bar{d}_{2}})\bar{p}\bar{q}\bar{p}_{r}\bar{q}_{r},\nonumber\\
\rho_{77} &=& \frac{1}{\mathcal{N}}|\gamma|^{2}\bar{\mu}\bar{d}_{2}d_{2}\bar{p}_{r}\bar{q}_{r}^{2}\bar{q}^{2},\nonumber\\
\rho_{99} &=& \frac{1}{\mathcal{N}} |\gamma|^{2}(\bar{\mu}\bar{d}^{2}_{2}+\mu \bar{d}_{2})\bar{q}_{r}^{2}\bar{q}^{2}.
\end{eqnarray}
where $\overline{\square}=1-\square$. And $\mathcal{N} = \alpha^{2}\bar{p}_{r}^{2}\bar{q}_{r}^{2}
+|\beta|^{2}\bar{p}^{2}\bar{p}_{r}^{2}[1+\bar{\mu} d_{1}^{2}q_{r}^{2}-2\bar{\mu}d_{1}q_{r}-\mu d_{1}(2q_{r}-q_{r}^{2})]
+|\gamma|^{2}\bar{q}^{2}\bar{q}_{r}^{2}[1+\bar{\mu} d_{2}^{2}p_{r}^{2}-2\bar{\mu}d_{2}p_{r}-\mu d_{2}(2p_{r}-p_{r}^{2})]$.

In order to achieve the optimal entanglement protection, it is essential to identify the optimal QMR parameters. While this can be accomplished by traversing $p_{r}$ and $q_{r}$ between 0 and 1, the optimal parameters obtained by this method may vary based on the specific initial state parameters and the correlation factor. Given the difficulty in precisely defining the parameters of the initial state and the correlation factor, this method is not a practical solution in the field of physics. In light of Fig. \ref{fig1}, it can be observed that the decays of both uncorrelated AD noise and FCAD noise can be regarded as quantum transitions that result in the excited states jumping to the ground state with varying probabilities. Consequently, it is possible to utilize the technique of `unravelling' the excitation into `jump' and `no-jump' scenarios, which allows for the manipulation of pure states \cite{scully_1997_quantum}. The optimal strength of the QMRs yields to
 \begin{equation}
 \label{eq13}
{p}_{r,\rm WM}^{\rm opt}=1-\bar{q}\bar{d}_{2},\ \ \ \ {q}_{r,\rm WM}^{\rm opt}=1-\bar{p}\bar{d}_{1},
\end{equation}
Although Eq. (\ref{eq13}) may not be the most optimal result, it is not dependent on any specific parameters of the initial state, and thus remains valid for any initial state. Accordingly, the optimality condition allows us to determine the entanglement $N_{1,\rm WM}^{\rm opt}$ and $N_{2,\rm WM}^{\rm opt}$ for the initial states $|\psi\rangle_{1}$ and $|\psi\rangle_{2}$ respectively.
 

\begin{figure}
\begin{center}
\includegraphics[width=16cm] {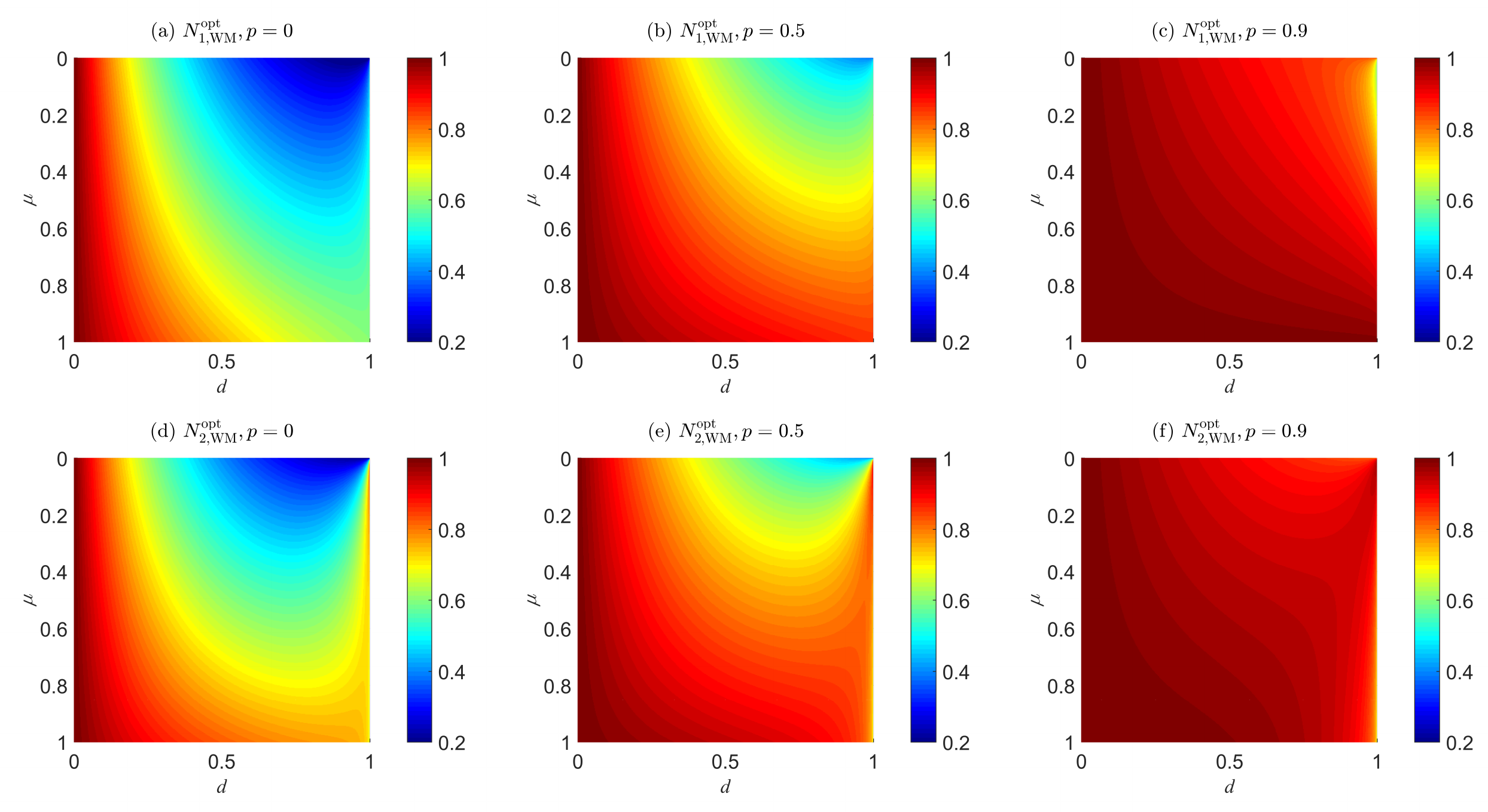}
\caption{(color online) (a)-(c) $N_{1,\rm WM}^{\rm opt}$ and (d)-(f) $N_{2,\rm WM}^{\rm opt}$ as a function of  $d$ and $\mu$ for the optimal strength of QMR of Eq. (\ref{eq13}). The remaining parameters are the same as Fig. \ref{fig2}.}
\label{fig3}
\end{center}
\end{figure}

Considering that both WM and QMR are non-unitary operations, it can be seen that this entanglement-protecting scheme is actually probabilistic. Since the CAD channel is trace-preserving, the successful probability is not affected by this property. The probability of success is therefore determined by the success of WM and QMR in sequence.
In accordance with the standard postulate of quantum measurement, the successful probability can be obtained in two steps.
Firstly, we calculate the probability of obtaining the measurement outcome of ${M}_{\rm WM}$, which is $\mathcal{P}_{1}=\text{tr}\left[M_{\rm WM}^{\dagger}M_{\rm WM}\rho(0)\right]$. Secondly, we calculate the probability of obtaining the measurement outcome of QMR, which is $\mathcal{P}_{2}=\text{tr}\left[M_{\rm R}^{\dagger}M_{\rm R}\mathcal{E}_{\rm CAD}\left[\rho(0)\right]\right]$.
The final probability of this scheme is dependent on the successful performance of both the WM and QMR operations in sequence. This can be expressed as $P_{\rm WM}=\mathcal{P}_{1}\mathcal{P}_{2}$, which is equivalent to the normalization factor
$\mathcal{N}$. 
\begin{equation}
\label{eq14}
P_{1,\rm WM}=\alpha^{2}\bar{p}_{r}^{2}\bar{q}_{r}^{2}
+|\beta|^{2}\bar{p}^{2}\bar{p}_{r}^{2}\left[1+\bar{\mu}d_{1}^{2}q_{r}^{2}-2\bar{\mu}d_{1}q_{r}-\mu d_{1}(2q_{r}-q_{r}^{2})\right]
+|\gamma|^{2}\bar{q}^{2}\bar{q}_{r}^{2}\left[1+\bar{\mu} d_{2}^{2}p_{r}^{2}-2\bar{\mu}d_{2}p_{r}-\mu d_{2}(2p_{r}-p_{r}^{2})\right].
\end{equation}
The probability of success for $N_{1,\rm opt}^{\rm WM}$ and $N_{2,\rm opt}^{\rm WM}$ can be calculated under the condition of Eq. (\ref{eq13}), which are denoted as $P_{1,\rm WM}^{\rm opt}$ and $P_{2,\rm WM}^{\rm opt}$.

\begin{figure}
\begin{center}
\includegraphics[width=16cm] {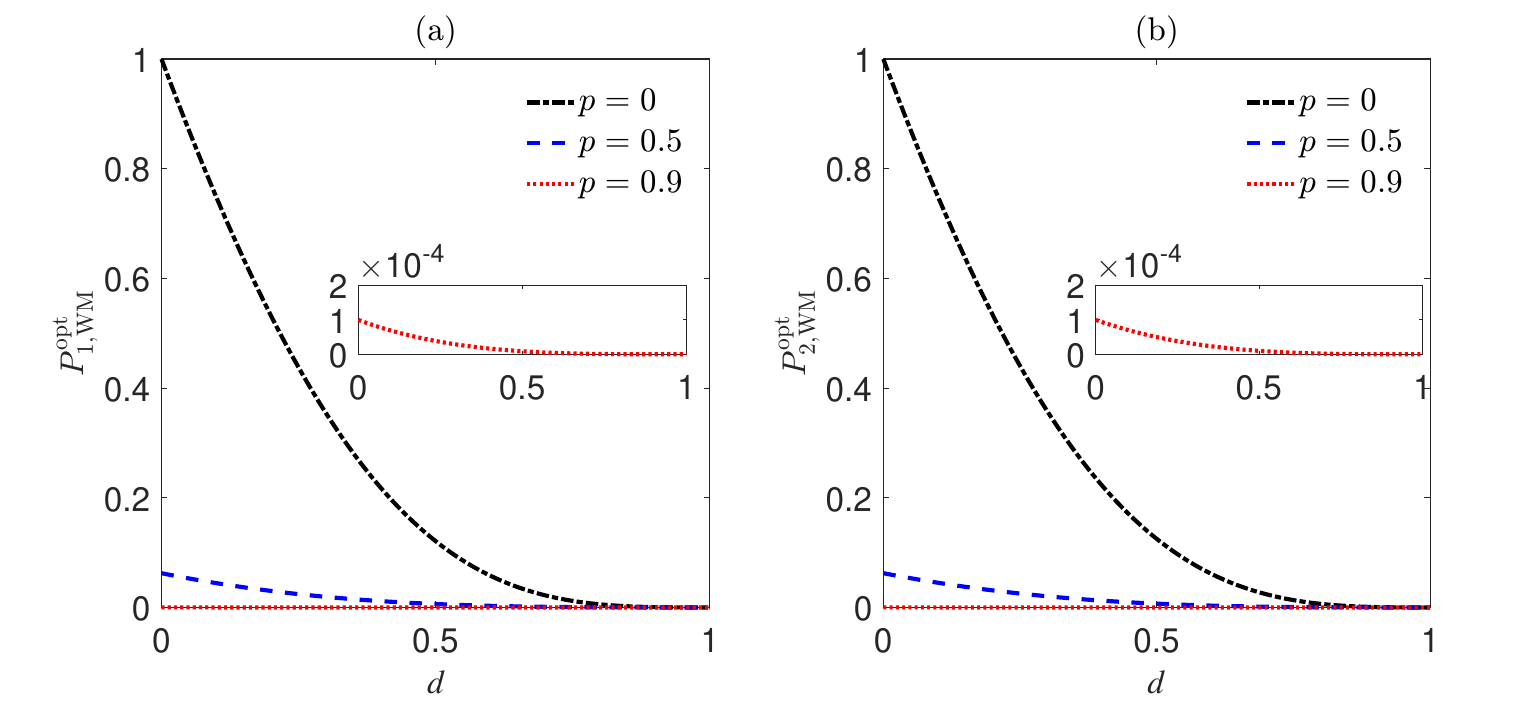}
\caption{(color online) (a) $P_{1,\rm opt}^{\rm WM}$ and (b) $P_{2,\rm opt}^{\rm WM}$ as a function of  $d$ for the optimal strength of QMR of Eq. (\ref{eq13}) with $\mu=0.6$. The insertions show the magnified plots of $P_{1,\rm WM}^{\rm opt}$ and $P_{2,\rm WM}^{\rm opt}$ for $p=0.9$. The remaining parameters are the same as Fig. \ref{fig2}.}
\label{fig4}
\end{center}
\end{figure}

We plot in Fig. \ref{fig3} the behaviors of $N_{1,\rm WM}^{\rm opt}$ and $N_{2,\rm WM}^{\rm opt}$ as a function of the noise parameter $d$ and the correlation parameter $\mu$ for different values of $p$. Figure \ref{fig4} shows the behaviors of successful probabilities. From Figs. \ref{fig3} and \ref{fig4}, we can draw the following conclusions:

(i) In contrast to Fig. \ref{fig2}, even when $p = 0$ (but $q\neq0$), the entanglement between the two qutrits does not disappear completely with increasing $d$. Conversely, a portion of the entanglement persists, with a value exceeding 0.2, as illustrated in Figs. \ref{fig3}(a) and  \ref{fig3}(d). 

(ii) The second significant observation is that the protection of entanglement is enhanced as $p$ increases. As $p$ approaches 1, a substantial portion of the initial entanglement can be recovered. Nevertheless, this is achieved at the expense of a low success probability, as illustrated in Fig. \ref{fig4}. It can be observed that the success probability decreases with increasing WM strength.

(iii) In the case that WM is introduced, the correlation effect may not always be favorable to the enhancement of entanglement, see Fig. \ref{fig2}(c). The reason for this is that during the reversal procedure, QMR is unable to exactly distinguish between the uncorrelated transitions, such as $|11\rangle\rightarrow|10\rangle,|01\rangle\rightarrow|00\rangle$ (or $|22\rangle\rightarrow|20\rangle,|02\rangle\rightarrow|00\rangle$), and fully correlated transitions, such as $|11\rangle\rightarrow|00\rangle$ (or $|22\rangle\rightarrow|00\rangle$). Consequently, the entanglement could not be restored to its original state through the application of QMR if both uncorrelated and fully correlated transitions were present simultaneously, as illustrated in Figs. \ref{fig3}(c) and  \ref{fig3}(f). Nevertheless, the WM scheme works well in the uncorrelated AD channel and FCAD channel.

One might inquire as to whether the observed result is attributable to the fact that condition (\ref{eq13}) is not the most optimal. It can be stated with a high degree of certainty that full restoration of entanglement cannot be achieved even by traversing all possible values for $p_{r}$ and $q_{r}$ when $\mu\neq0,1$. We conjecture that a potentially possible strategy would be to redesign the form of the QMR in a manner that would enable it to distinguish between the two transition mechanisms of uncorrelated AD and FCAD channels (e.g., this could be achieved by discriminating the number of radiated photons in real time). However, such an operation must be performed on both quantum systems simultaneously, and thus represents a non-local operation that is beyond the scope of this paper.

%

\section{Protecting qutrit-qutrit entanglement by EAM and QMR}
\label{sec5}
We now turn to discuss the protection of entanglement through EAM and QMR. The EAM scheme is analogous to an error correction scheme. The distinction between WM and EAM is evident from the perspective that WM is a preemptive operation that is applied to the quantum system itself, whereas EAM is a postemptive monitoring process that is performed on the environment. The determination of whether a dissipative jump has occurred in the quantum system is based on the observation of a change in excitons in the environment. In the case that the detector does not register a click (i.e., no-click), this signifies that the quantum system has not undergone a dissipative transition. In such a scenario, the performance of a corrective operation on the quantum system can restore the entanglement of the quantum system. As demonstrated by Eqs. (\ref{eq2}) and (\ref{eq4}), we find that only the elements $E_{00}$ and $A_{00}$ are incapable of inducing dissipative transitions within the quantum system. Consequently, the complete process can be described by the map
\begin{equation}
\rho_{\rm{EAM}}= \frac{1}{\mathcal{N}'}M_{\rm R}\left[(1-\mu)E_{00}\rho(0)E_{00}^{\dagger}+\mu A_{00}\rho(0)A_{00}^{\dagger}\right]M_{\rm R}^{\dagger}.
\end{equation}

The non-zero elements of the normalized $\rho_{1,\rm{EAM}}$ are
\begin{eqnarray}
\rho_{11}&=& \frac{1}{\mathcal{N}'}\bar{p}_{r}^{2}\bar{q}_{r}^{2}(\frac{\bar{\mu}}{G_{1}}+\frac{\mu}{G_{2}})\alpha^{2},\nonumber\\
\rho_{15}&=& \rho_{51}^{*} = \frac{1}{\mathcal{N}'}\bar{p}_{r}^2\bar{q}_{r}(\frac{\bar{\mu}\bar{d}_{1}}{G_{1}}+\frac{\mu\sqrt{\bar{d}_{1}}}{G_{2}})\alpha\beta^{*},\nonumber\\
\rho_{19}&=& \rho_{91}^{*} = \frac{1}{\mathcal{N}'}\bar{p}_{r}\bar{q}_{r}^{2}(\frac{\bar{\mu}\bar{d}_{2}}{G_{1}}+\frac{\mu\sqrt{\bar{d}_{2}}}{G_{2}})\alpha\gamma^{*},\nonumber\\
\rho_{55}&=& \frac{1}{\mathcal{N}'}\bar{p}_{r}^{2}(\frac{\bar{\mu}\bar{d}_{1}^{2}}{G_{1}}+\frac{\mu \bar{d}_{1}}{G_{2}})|\beta|^{2},\nonumber\\
\rho_{59}&=& \rho_{95}^{*} = \frac{1}{\mathcal{N}'}\bar{p}_{r}\bar{q}_{r}(\frac{\bar{\mu}\bar{d}_{1}\bar{d}_{2}}{G_{1}}+\frac{\mu\sqrt{\bar{d}_{1}\bar{d}_{2}}}{G_{2}})\beta\gamma^{*},\nonumber\\
\rho_{99}&=& \frac{1}{\mathcal{N}'}\bar{q}_{r}^{2}(\frac{\bar{\mu}\bar{d}_{2}^{2}}{G_{1}}+\frac{\mu \bar{d}_{2}}{G_{2}})|\gamma|^{2}.
\end{eqnarray}
where $G_{1}=\alpha^{2}+|\beta|^{2}\bar{d}_{1}^{2}+|\gamma|^{2}\bar{d}_{2}^{2}$, $G_{2}=\alpha^{2}+ |\beta|^{2}\bar{d}_{1}+|\gamma|^{2}\bar{d}_{2}$ and
$\mathcal{N}'=\frac{1}{G_{1}G_{2}}[\bar{p}_{r}^{2}\bar{q}_{r}^{2}(\mu G_{1}+\bar{\mu}G_{2})\alpha^{2}+\bar{p}_{r}^{2}(\mu d_{1} G_{1}+\bar{\mu}\bar{d}_{1}^{2} G_{2})|\beta|^{2}+\bar{q}_{r}^{2}(\mu d_{2} G_{1}+\bar{\mu}\bar{d}_{2}^{2} G_{2})|\gamma|^{2}]$.

\begin{figure}
\begin{center}
\includegraphics[width=16cm] {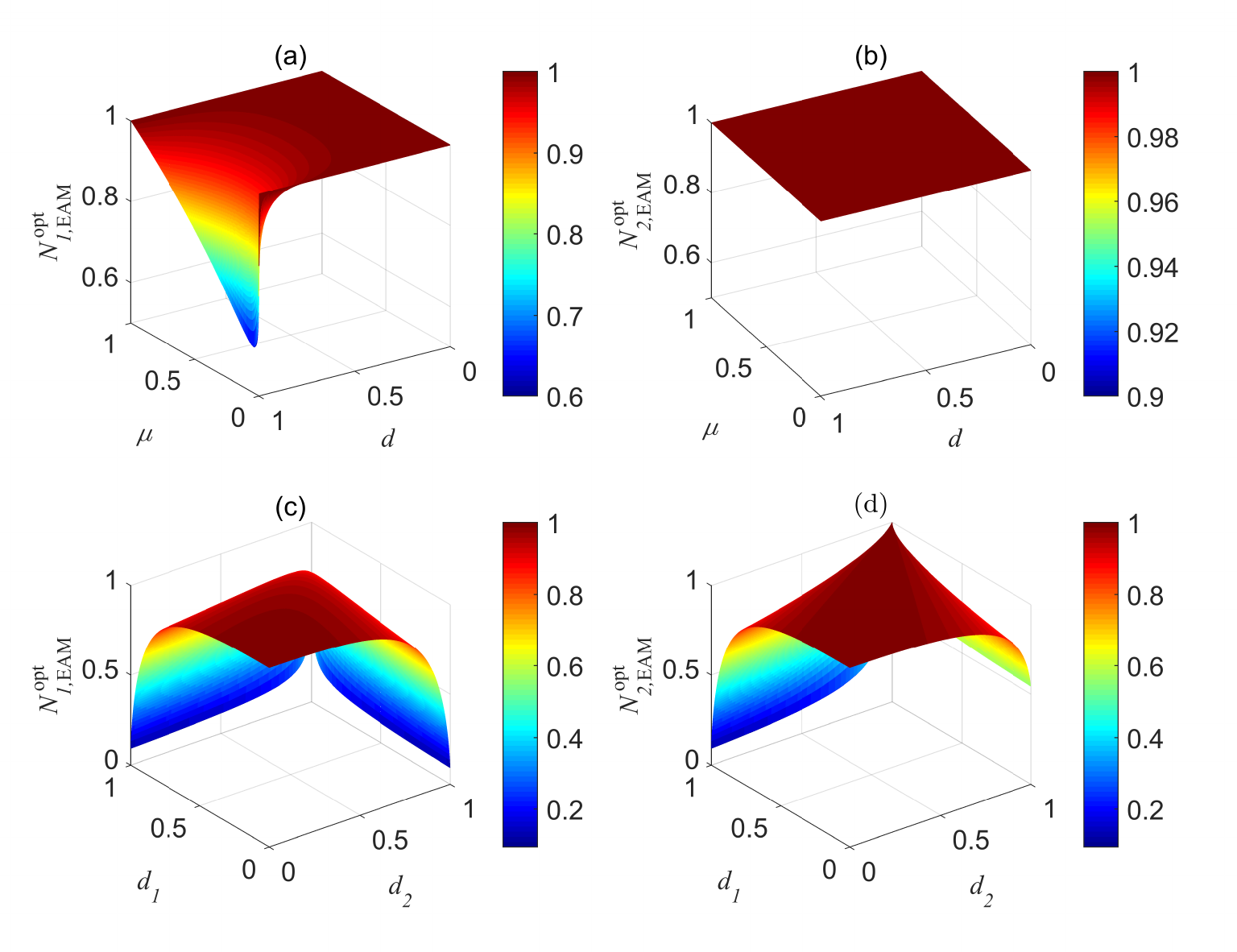}
\caption{(color online) (a) $N_{1,\rm EAM}^{\rm opt}$ and (b) $N_{2,\rm EAM}^{\rm opt}$ as a function of $d$ and $\mu$. (c) $N_{1,\rm EAM}^{\rm opt}$ and (d) $N_{2,\rm EAM}^{\rm opt}$ as a function of $d_1$ and $d_{2}$ with $\mu=0.6$. The remaining parameters are the same as Fig. \ref{fig2}.}
\label{fig5}
\end{center}
\end{figure}

In light of the aforementioned considerations pertaining to the WM scheme, the optimal strength of QMR is determined to be
\begin{equation}
\label{eq17}
p_{r,\rm EAM}^{\rm opt}=d_{1},\ \ \ q_{r,\rm EAM}^{\rm opt}= d_{2}.
\end{equation}
The successful probability is also given by the normalization factor 
 \begin{equation}
\label{eq18}
P_{1,\rm EAM}=\frac{1}{G_{1}G_{2}}\left[\bar{p}_{r}^{2}\bar{q}_{r}^{2}(\mu G_{1}+\bar{\mu}G_{2})\alpha^{2}+\bar{p}_{r}^{2}(\mu d_{1} G_{1}+\bar{\mu}\bar{d}_{1}^{2} G_{2})|\beta|^{2}+\bar{q}_{r}^{2}(\mu d_{2} G_{1}+\bar{\mu}\bar{d}_{2}^{2} G_{2})|\gamma|^{2}\right].
\end{equation}

Figures \ref{fig5}(a) and \ref{fig5}(b) illustrate the curves of $N_{1,\rm{EAM}}^{\rm opt}$ and  $N_{2,\rm{EAM}}^{\rm opt}$ as a function of $d$ and $\mu$. It is found that the initial entanglement of state $|\psi\rangle_{2}$ has been restored with the assistance of EAM and QMR. Similarly, it is observed that the entanglement does not monotonically increase with the correlation factor $\mu$ once the parameter $d$ has been determined. For example, the behavior of  $N_{1,\rm{EAM}}^{\rm opt}$ when $d\rightarrow1$, as shown in Fig. \ref{fig5}(a). A straightforward interpretation is that during the action of EAM, we select only those results that the detector does not respond to (i.e., no-click), implying that the system is influenced by only two evolutionary processes, $E_{00}$ and $A_{00}$. However, subsequent QMR operations cannot strictly distinguish the relative weights of these two processes. It appears that the correlation effect does not contribute to the preservation of entanglement in this context. 


We note the entanglement protection of state $|\psi\rangle_{2}$ is satisfactory in Fig. \ref{fig5}(b). It is advantageous to set $d_{1} = d_{2}$, as this ensures that the state after QMR is perfectly recovered to the initial state. If $d_{1}\neq d_{2}$, the initial entanglement will be not fully restored, but $N_{2,\rm{EAM}}^{\rm opt}$ is still superior to $N_{1,\rm{EAM}}^{\rm opt}$, as shown in Figs. \ref{fig5}(c) and \ref{fig5}(d).
This is due to the fact that only one of the states, namely $|11\rangle$ in $|\psi\rangle_{2}$, will be affected by the FACD and the uncorrelated AD noise simultaneously, whereas the other two states, namely $|02\rangle$ and $|20\rangle$, will only experience the effect of the uncorrelated AD noise. It is therefore possible to protect the entanglement between the states $|02\rangle$ and $|20\rangle$ by selecting an appropriate QMR dedicated to eliminating the effects of AD noise.

\begin{figure}
\begin{center}
\includegraphics[width=16cm] {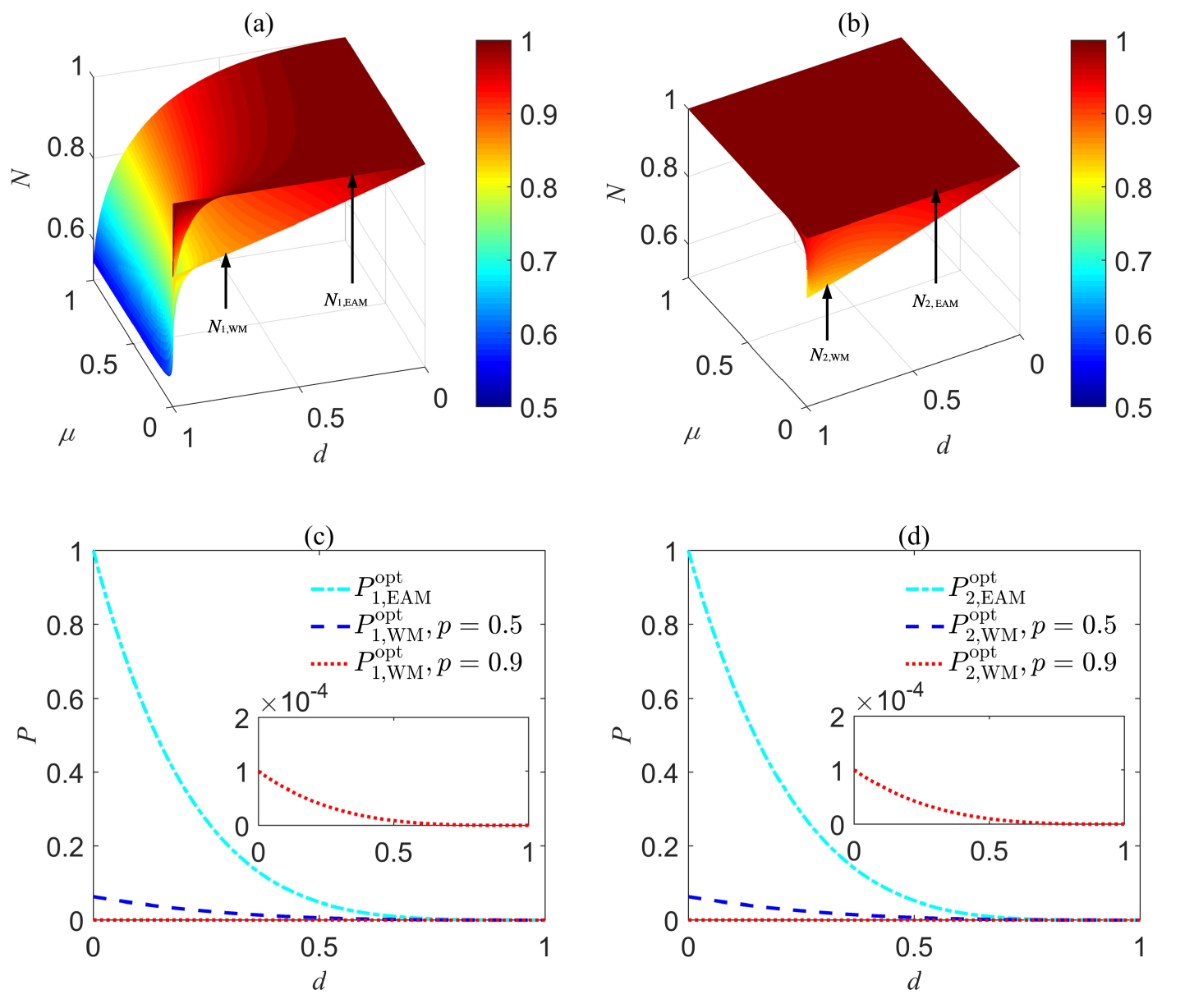}
\caption{(color online) Comparison of the entanglement and the success probability between the EAM scheme and the WM scheme. (a) $N_{1,\rm EAM}^{\rm opt}$ and $N_{1,\rm WM}^{\rm opt}$ as a function of $d$ and $\mu$. (b) $N_{2,\rm EAM}^{\rm opt}$ and $N_{2,\rm WM}^{\rm opt}$ as a function of $d$ and $\mu$.  (c) $P_{1,\rm EAM}^{\rm opt}$ and $P_{1,\rm WM}^{\rm opt}$ as a function of $d$ with $\mu=0.6$. (d) $P_{2,\rm EAM}^{\rm opt}$ and $P_{2,\rm WM}^{\rm opt}$ as a function of $d$ with $\mu=0.6$. The insertions in (c) and (d) show the magnified plots of $P_{1,\rm WM}^{\rm opt}$ and $P_{2,\rm WM}^{\rm opt}$ for $p=0.9$. The remaining parameters are the same as Fig. \ref{fig2}.}
\label{fig6}
\end{center}
\end{figure}

We conclude this section with a brief comparison between the EAM scheme and the WM scheme. 
As illustrated in Fig. \ref{fig6}, even when $p$ is set to 0.9 in the WM scheme, the degree of entanglement achieved remains below that of the EAM scheme. With regard to the success probability, the EAM scheme also exhibits a higher level of performance than the WM scheme. This indicates that the EAM scheme performs better than the WM scheme. The reason can be attributed solely to the differences between WM and EAM. As WM is performed prior to the quantum system passes through the CAD channel, it can only obtain part of the information about the quantum system but not about the channel. On the contrary, EAM is performed after the quantum system has passed through the CAD channel, thereby facilitating the acquisition of information about both the quantum system and the channel. Therefore, in the subsequent correction operation, a more suitable QMR can be selected to enhance the protection of the entanglement.

%
%
%

\section{Discussions}
\label{sec6}
It is essential to provide a concise overview of the principal techniques employed in the experimental implementation of our procedures. In this context, we will limit our discussion to the cavity quantum electrodynamics (QED) system, which we believe is the most promising candidate for the experimental realization of our schemes.

\textbf{Initial state preparation.} The qutrit-qutrit entanglement of Eqs.~(\ref{eq6}) and (7) can be generated by sending a pair of momentum and polarization-entangled photons to two spatially separated cavities in which a $V$-type atom is trapped, as described in Ref.~\cite{lloyd_2001_long}. For illustrative purposes, we consider the energy structure of the $^{87}\rm{Rb}$ atom. 
The state $|0\rangle$ corresponds to the atomic level $|F=1, m_{F}=0\rangle$ of $5^{2}S_{1/2}$, while the states $|1\rangle$ and $|2\rangle$ correspond to the degenerate levels $|F=1, m_{F}=1\rangle$ and $|F=1, m_{F}=-1\rangle$ of $5^{2}P_{1/2}$, respectively.  The transitions $|1\rangle\rightarrow|0\rangle$ and $|2\rangle\rightarrow|0\rangle$ emit right- and left-circularly polarized photons, respectively.

\textbf{CAD noise.} In the cavity QED system, the spontaneous radiation process of the atom can be viewed as a form of AD noise. Given that the two excited states of the aforementioned three-energy system are degenerate, it follows that only the transitions $|1\rangle\rightarrow|0\rangle$ and $|2\rangle\rightarrow|0\rangle$ exist. This spontaneous radiation process can be described by Eq. (\ref{eq1}) \cite{chciska_2007_separability}. When two atoms pass through the cavity consecutively at short intervals, there is a certain probability of correlated radiation, namely $|11\rangle\rightarrow|00\rangle$ or $|22\rangle\rightarrow|00\rangle$. The correlated radiation process will be described by Eq. (\ref{eq3}). The total decoherence process encompasses both individual AD noise and FCAD noise, which can be described in terms of Eq. (\ref{eq5}), where $\mu$ represents the associated probability of fully correlated radiation.

\textbf{WM.} The WM of $E_{\rm WM}$ is also referred to as a null-result measurement. This is distinct from a conventional projective measurement, which is typically employed to extract information accompanied by the collapse of the quantum system. 
The realization of WM hinges on the absence of a collapsing projection of the quantum system, which is to say that no signal is detected. WM can be implemented with different technologies on different platforms for the qubit system \cite{katz_2008_reversal,kim_2011_protecting}. For the qutrit system, a complete set of POVM operators can be constructed that contains $E_{\rm WM}={\rm diag}(1,\sqrt{\bar{p}},\sqrt{\bar{q}})$, $E_{2}={\rm diag}(0,\sqrt{p},0)$ and $E_{3}={\rm diag}(0,0,\sqrt{q})$. It should be noted that the measurement operators $E_{2}$ and $E_{3}$ result in the collapse of the quantum state to the ground state, while $E_{\rm WM}$ keeps the quantum system to be a coherent superposition state. 
In the cavity QED system, one can infer whether an atom is in a coherent superposition state by measuring the emitted spectrum. If the atom has emitted a photon (it should be noted that the right-circularly polarized photon and the left-circularly polarized photon correspond to transitions $|1\rangle\rightarrow|0\rangle$ and $|2\rangle\rightarrow|0\rangle$, respectively), it is not in an excited state. In the case that a photon is detected (regardless of right- or left-circularly polarized photon), the result is discarded. Conversely, if no photon is detected, the result is retained. Such a no click result represents the implementation of $E_{\rm WM}$.
The parameters $p$ and $q$ can be determined by counting the right- and left-circularly polarized photons.


\textbf{QMR.} In order to see clearly how QMR $E_{\rm R}$  is realized, we re-express QMR in the following form
\begin{equation}
E_{\rm R}=\left(
        \begin{array}{ccc}
         \sqrt{(1-p_{r})(1-q_{r})} & 0
          & 0 \\
          0 & \sqrt{1-q_{r}} & 0 \\
          0 & 0 & \sqrt{1-p_{r}} \\
        \end{array}
      \right)=\sqrt{\bar{p}_{r}\bar{q}_{r}}\left(E_{\rm WM}\right)^{-1}=\sqrt{\bar{p}_{r}\bar{q}_{r}}{T}E_{\rm WM}{T}E_{\rm WM}{T},
\end{equation}
where ${T}=|0\rangle\langle2|+|1\rangle\langle0|+|2\rangle\langle1|$ is the trit-flip operation.
Thus, the procedure of QMR can be constructed by the following five sequential operations on each
system qutrit: trit-flip (${T}$), weak measurement ($E_{\rm WM}$), trit-flip (${T}$), another weak measurement
($E_{\rm WM}$), and trit-flip (${T}$). The trit-flip operation ${T}$ can be realized by a $\pi$ pulse applied on the
transition $|1\rangle\leftrightarrow|2\rangle$ and followed by another $\pi$ pulse to interchange the populations between $|0\rangle$ and $|1\rangle$. (i.e., by the series of two $\pi$ pulses $\pi^{|1\rangle\leftrightarrow|2\rangle}\pi^{|0\rangle\leftrightarrow|1\rangle}$)
\cite{das_2003_quantum}.

\textbf{EAM.} Unlike WM, which acts on the quantum system itself, EAM acts on the environment coupled to the quantum system. 
All we need to do is add a detector to monitor the photon changes in the environment. If the environment is initially in a vacuum state and the detector does not register any photon production, then the EAM is realized.

\section{Conclusions}
\label{sec7}
In conclusion, two schemes have been proposed to protect the qutrit-qutrit entanglement in the CAD channel.
The WM scheme has been designed to prevent the complete collapse of the quantum state. When combined with QMR, it is capable of eliminating the effects of CAD noise and improving entanglement. It has been demonstrated that while the protection of entanglement can be enhanced by increasing the WM strength, this improvement is accompanied by a reduction in the probability of success. The EAM approach involves monitoring the channel and then performing a QMR operation on the system based on the measurement outcomes. The results demonstrate that the EAM+QMR approach can almost restore the initial entanglement state, effectively counteracting the decoherence effects of CAD noise. Furthermore, the EAM method has been demonstrated to be a more effective approach for protecting entanglement, with higher success probabilities compared to the WM method. This advantage arises from the fact that EAM itself is a backward operation that collects information about the quantum system and the channel simultaneously. This enables the design of a more effective QMR, thereby enhancing the protection of entanglement. Finally, we have discussed how the aforementioned two schemes can be realized based on the cavity QED system.

Our findings are of significant consequence for the development of robust quantum information protocols in the NISQ era, where the quality of quantum entanglement is of paramount importance. Although both schemes are probabilistic, the relative importance of achieving a high probability of success versus the amount of quantum entanglement may vary in practical quantum information processing tasks. In certain instances, attaining a high probability of success may be more crucial (e.g., in quantum computation), while in other cases, the degree of quantum entanglement may be more significant (e.g., in quantum parameter estimation). Consequently, the trade-off between the substantial retrieved concurrence and the low success probability must be carefully considered in a realistic context.

\section*{CRediT authorship contribution statement}
\textbf{X.~Xiao:} main contributor, analytical calculations and numerical simulations, visualization,
\textbf{W.~R.~Huang:} numerical simulations, 
\textbf{T.~X.~Lu:} conducted numerical calculations, 
\textbf{Y.~L.~Li:} designed and supervised the research.
All authors reviewed and edited the manuscript.

\section*{Declaration of Interests}
The authors declare that they have no known competing financial interests or personal relationships that could have appeared to influence the work reported in this paper.


\section*{Acknowledgments}
This work was supported by the Funds of the National Natural Science Foundation of China under Grant Nos. 12265004, 12365003, 12205054 and Jiangxi Provincial Natural Science Foundation, China under Grant Nos. 20212ACB211004.

%
%

\bibliographystyle{elsarticle-num}

\bibliography{refs}

\vskip3pt
\end{document}